\documentclass[preprint,showpacs,preprintnumbers,amsmath,amssymb,superscriptaddress]{revtex4}

\usepackage[dvipdfm]{graphicx}
\usepackage{dcolumn}
\usepackage{bm}
\usepackage{amsmath, subfigure}

\begin{document}

\preprint{}

\title{Sparse and Dense Encoding in Layered Associative Network of Spiking Neurons}

\author{Kazuya Ishibashi}
 \email{kazuya@mns.k.u-tokyo.ac.jp}
\affiliation{Graduate School of Frontier Sciences, The University of Tokyo, Japan}
\author{Kosuke Hamaguchi}%
\affiliation{Brain Science Institute, RIKEN}
\author{Masato Okada}
\affiliation{Graduate School of Frontier Sciences, The University of Tokyo, Japan}
\affiliation{Brain Science Institute, RIKEN}

\date{\today}

\begin{abstract}
A synfire chain is a simple neural network model which can propagate stable synchronous spikes called a pulse packet and widely researched.  However how synfire chains coexist in one network remains to be elucidated.  We have studied the activity of a layered associative network of Leaky Integrate-and-Fire neurons in which connection we embed memory patterns by the Hebbian Learning.  We analyzed their activity by the Fokker-Planck method.  In our previous report, when a half of neurons belongs to each memory pattern (memory pattern rate $F=0.5$), the temporal profiles of the network activity is split into temporally clustered groups called sublattices under certain input conditions.  In this study, we show that when the network is sparsely connected ($F<0.5$), synchronous firings of the memory pattern are promoted.  On the contrary, the densely connected network ($F>0.5$) inhibit synchronous firings.  The sparseness and denseness also effect the basin of attraction and the storage capacity of the embedded memory patterns.  We show that the sparsely(densely) connected networks enlarge(shrink) the basion of attraction and increase(decrease) the storage capacity.
\end{abstract}

\pacs{Valid PACS appear here}
\maketitle

\section{Introduction}
What is the role of synchronous spikes?  It is an important and challenging question in neuroscience.  In order to approach this question, the ability to generate synchronous spikes in various networks have been studied.  A synfire chain~\cite{corticonics} is a functional feed-forward network and able to transmit synchronous spikes called a pulse packet.  Some experimental results imply the existence of synfire chains \textit{in vivo}~\cite{abeles93, abeles98} and \textit{in vitro}~\cite{Yuste03, ikegaya}.  Synfire chains have been also intensively studied theoretically,~\cite{diesmann,gewaltig,cateau, rossum, gerstner, hamaguchi, hamaguchiBC, hamaguchiNC, aviel, aviel2003} and have been confirmed to exist \textit{in vitro} in an iteratively constructed network~\cite{reyes}.

Although Abeles has already mentioned the idea of embedding multiple synfire chains in a local network, many of the studies on synfire chains use a homogeneously connected network~\cite{diesmann, gewaltig, cateau}.   One way to embed multiple synfire chains is to sum up the synaptic connections of synfire chains like an associative memory network.  Even if each synfire chain consists of homogeneous connections, the connections summed them up are inhomogeneous.  The ability to generate synchronous spikes in homogeneous network has been enthusiastically studied, but that in inhomogeneous network have not been throughly investigated.  Here we embed the chains in the way of associative memory.  Therefore we pay attention to whether the network transmits synchronous firings of the embedded memory patterns.

We have previously reported the activity of a network in which the half of neurons join in each memory pattern~\cite{ishibashi}.  In this paper we analyze sparsely and densely connected networks and discuss the difference of ability to generate synchronous spikes among those networks.  The sparsely connected networks of the associative memory constructed by the binary neurons have been intensively researched~\cite{amari89, amari91, okada}.  However, the dynamics of layered associative networks with spiking neurons are not throughly studied.

Section 2 explains the details of our layered associative network, and \S 3 explains the Fokker-Planck method.  Section 4 describes the result of our analysis.  In \S \ref{sec_single}, we show the activity in single pattern activation.  In \S \ref{sec_double} we address the activity in two patterns activation.  More specifically, we studied the network activity in response to the activation of two patterns with different strength but with the same timing (\S \ref{sec_diffstr}), and with the same strength but with different timing (\S \ref{sec_difftime}).  The basin of attraction is also studied (\S \ref{sec_attract}).  Section \ref{sec_cap} is devoted for the memory capacity of the network.  Section 5 is a summary and discussion.
\section{Layered Associative Network}
In this paper, we consider a layered associative memory network with the standard Hebbian connections~\cite{hopfield, cooper, amit, amitbook}.  We constructed feedforward associative networks by using the conventional method of sparsely connected network~\cite{amari89, amari91, okada, domany}.  Here a synaptic connection $J_{ij}^l$ from the $i$th neuron on layer $l$ to the $j$th neuron on layer $l+1$ is given by
\begin{align}
J^l_{ij} = \frac{1}{F(1-F)N} \sum^p_{\mu=1} (\xi^{l+1, \mu}_j -F)(\xi^{l, \mu}_i-F),\label{eq_hebb}
\end{align}
where the index of a neuron in a layer is $i, j = 1, \ldots, N$, and the index of the a memory pattern is $\mu=1,\ldots,p$, and $\xi^{l,\mu}_i$ represents the $\mu$th embedded memory pattern of the $i$th neuron on layer $l$ and takes on a value of either $+1$ or $0$ according to the following probability,
\begin{align}
\mathrm{Prob.}[\xi^{l,\mu}_i = +1] = F.
\end{align}
We call the probability of $\xi^{l,\mu}_i=+1$ the 'pattern rate' $F$.  Then eq. (\ref{eq_hebb}) satisfies that the expected value of sum of synaptic connection $\sum_{i,j}J^l_{ij}$ equals 0, which means that excitation and inhibition are balanced regardless of pattern rate $F$.

\begin{figure}[tb]
\centering
\includegraphics[width=6cm]{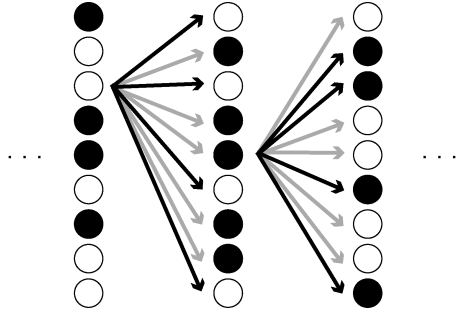}
\caption{Schematic diagram of our layered associative network.}
\label{fig_intro}
\end{figure}

A memory pattern $\xi^{l,\mu}_i=+1$ means that the neuron should fire in memory pattern $\mu$ and $\xi^{l,\mu}_i=0$ means the neuron should be silent.  We analyze the network activity when we change the pattern rate $F$.  Figure~\ref{fig_intro} is a schematic diagram of this network.

We use the leaky integrate-and-fire (LIF) neuron model, and the dynamics of the membrane potential $v^l_i(t)$ can be described as a stochastic differential equation,
\begin{align}
\frac{{\rm d} v^l_i(t)}{{\rm d} t} = - \frac{v^l_i(t)-V_{\mathrm{rest}}}{\tau} + \frac{I^{l, \alpha}_i(t) + I_0}{C} + D' \eta(t),\label{eq_lif}
\end{align}
where $\tau$ is the membrane time constant, $V_{\mathrm{rest}}$ is the resting potential, $I_0$ is the mean of noisy input, $\eta(t)$ is white Gaussian noise satisfying $<\eta(t)>=0$ and $<\eta(t)\eta(t')>=\delta(t-t')$, and $D'$ is the amplitude of the noise.  Input current $I^{l,\alpha}_i(t)$ is obtained by convoluting the presynaptic firing $I^l_i(t)$ with the $\alpha$ function $\alpha(t) = \alpha^2 t \exp (-\alpha t)$ as follows.  $\beta$ is a conversion constant.
\begin{align}
I^{l,\alpha}_i(t) &= \beta \int^{\infty}_0 dt' \, \alpha(t')I^l_i(t-t'). \label{eq_ia}
\end{align}
Input current $I^l_i(t)$ is derived from the sum of synaptic connections in which neurons fire.  To equalize excitatory synaptic inputs for different pattern rate $F$, we set $I^l_i(t)$ proportional to $1/(1-F)$.
\begin{align}
I^l_i(t) &= \frac{1}{1-F} \sum^N_{j=1} J^{l-1}_{ji} \sum^n_{k = 1} \delta(t - t^{l-1}_{j, k}),
\end{align}
where $t^l_{i,k}$ indicates the times that the $i$th neuron on layer $l$ fires.  The $i$th neuron on layer $l$ has spiked $n$ times until time $t$.

Membrane potential dynamics follow the spike-and-reset rule; when the membrane potential $v^l_i(t)$ reaches the threshold $V_{\mathrm{th}}$, a spike is fired, and after the absolute refractoriness $t_{\mathrm{ref}}$, the membrane potential is reset to the resetting potential $V_{\mathrm{reset}}$.  By implementing the absolute refractoriness, burst firings are inhibited and we can focus on pulse packets propagation.

For the following analysis, we introduce the order parameter function $m^{l,\mu}(t)$, namely the overlap, defined by
\begin{align}
m^{l, \mu}(t) = \frac{1}{F(1-F)N} \sum^N_{i=1} (\xi^{l, \mu}_i-F) \sum^n_{k = 1} \delta(t - t^l_{i, k}).\label{eq_m}
\end{align}
Here, the overlap means how much the firing pattern matches the $\mu$th memory pattern on layer $l$.  If neurons with their memory patterns $\xi^{l,\mu}_i=+1$ fire once, then $\int^{\infty}_{-\infty}dt \, m^{l,\mu}(t)=1$.  By using the overlap, $I^l_i(t)$ can be rewritten as
\begin{align}
I^{l}_i(t) = \sum^p_{\mu=1} \frac{\xi^{l, \mu}_i -F}{1-F} m^{l-1, \mu}(t).\label{eq_i}
\end{align}
This means that the synaptic current to a neuron depends only on the overlap of the preceding layer and its memory patterns.  Here we can see that $I^l_i(t)$ need to be proportional to $1/(1-F)$ so that the excitatory inputs does not change with different pattern rate $F$.

Throughout this paper, the parameter values are fixed as follows: $V_{\mathrm{rest}}=V_{\mathrm{reset}}=0$ mV, $V_{\mathrm{th}}=15$ mV, $t_{\mathrm{ref}}=1$ ms, $\tau=10$ ms, $I_0=0.075$ pA, $C=100$ pF, $D'=1$, $\alpha=2$ ms$^{-1}$, and $\beta=0.17$ pA.  The number of neurons per layer is set to $N=5000$ in whole the LIF simulations.
\section{Fokker-Planck Method}
In this section, we introduce the analytical method of calculating the membrane potential distribution.  First, we define a vector whose elements are memory patterns of the $i$th neuron as $\bm{\xi^l_i} = ({\xi^{l,1}_i, \xi^{l,2}_i, \ldots ,\xi^{l,p}_i})$.  Each element takes on a value $+1$ or $0$, and thus this vector has $2^p$ combinations.  We can define $2^p$ groups according to $\bm{\xi}^l_i$ values.  We call each group a sublattice and we discriminate each sublattice with the vector $\bm{\xi} = ({\xi^{1}, \xi^{2}, \ldots ,\xi^{p}})$.  Each element $\xi^{\mu}$ takes on $+1$ or $0$ values.  Here we define $D(\bm{\xi})$ as the ratio of the number of neurons belonging to the sublattice $\bm{\xi}$ to the whole number of neurons and $D(\bm{\xi})$ is described as follows:
\begin{align}
D(\bm{\xi}) = \prod^p_{\mu=1} \left( \xi^\mu F + (1-\xi^\mu)(1-F) \right).
\end{align}

Neurons belonging to the same sublattice receive the same synaptic current, because the synaptic current depends on only the overlaps and its memory pattern $\bm{\xi}^l_i$ (eq. (\ref{eq_i})).  The distribution of the membrane potential is known to evolve according to the Fokker-Planck equation,~\cite{cateau, ishibashi}
\begin{align}
\frac{\partial}{\partial t}P^l_{\bm{\xi}}(v, t) &= -\frac{\partial}{\partial v}J^l_{\bm{\xi}}(v,t) + \delta(v-V_{\mathrm{reset}})D(\bm{\xi})\nu^l_{\bm{\xi}}(t-t_{\mathrm{ref}}),\label{eq_fp}
\end{align}
\begin{align}
J^l_{\bm{\xi}}(v,t) &= -\left( \frac{v-V_{\mathrm{rest}}}{\tau}-\frac{I^{l, \alpha}_{\bm{\xi}}(t)+\mu}{C}+\frac{\partial}{\partial v}\frac{{D'}^2}{2} \right) P^l_{\bm{\xi}}(v,t).\label{eq_j_fp}
\end{align}
$P^l_{\bm{\xi}}(v,t)$ is the distribution of the membrane potentials of the neurons belonging to sublattice $\bm{\xi}$, and $\nu^l_{\bm{\xi}}(t)=2^p J^l_{\bm{\xi}}(V_{\mathrm{th}},t)$ is the flow of probability across the threshold $V_{\mathrm{th}}$ per second per neuron, that is to say firing rate; the number of spikes per second per neuron.  $P^l_{\bm{\xi}}(v,t)$ and $\nu^l_{\bm{\xi}}(t)$ satisfy the normalization condition:
\begin{align}
\int^{V_{th}}_{-\infty}dv \, P^l_{\bm{\xi}}(v,t) + D(\bm{\xi})\int^{t_{\mathrm{ref}}}_{0} d \tau \, \nu^l_{\bm{\xi}}(t-\tau) =D(\bm{\xi}).
\end{align}

Here we define the overlap vector, $\bm{m}^l(t) = (m^{l,1}(t), m^{l,2}(t), \ldots, m^{l,p}(t) )$.  From eqs.~(\ref{eq_ia}) and (\ref{eq_i}), we can describe the synaptic current $I^{l, \alpha}_{\bm{\xi}}(t)$ by using $\bm{\xi}$ and $\bm{m}^l(t)$ as follows:
\begin{align}
I^{l, \alpha}_{\bm{\xi}}(t) &= \beta \int^{\infty}_0 dt' \, \alpha(t)I^l_{\bm{\xi}}(t-t'),\label{eq_ia_fp}\\
I^l_{\bm{\xi}}(t) &= \frac{\bm{\xi} -F \bm{I}}{1-F}\cdot \bm{m}^{l-1}(t) ,\label{eq_i_fp}
\end{align}
where $\bm{I}$ is the vector whose all elements are 1 and size is $p$.

From eq.~(\ref{eq_m}), we can describe the overlap $m^{l,\mu}(t)$ by using firing rate $\nu^l_{\xi}(t)$ as 
\begin{align}
m^{l,\mu}(t) = \frac{1}{F(1-F)} \sum_{\bm{\xi}} D(\bm{\xi}) \nu^l_{\bm{\xi}} (\xi^\mu (1-F)- (1-\xi^\mu)F) .\label{eq_m_fp}
\end{align}

Here we can describe the network dynamics only by using macroscopic parameters $P^l_{\bm{\xi}}(v,t)$, $\nu^l_{\bm{\xi}}(t)$, $\bm{m}^l(t)$, and $I^l_{\bm{\xi}}$.

In this paper, we numerically calculate the Fokker-Planck equation with the Chang-Cooper Method~\cite{fp, cc70, park}.  At the boundary, we stock the flow at $V_{\mathrm{th}}$ and add the stocked flow before $t_{\mathrm{ref}}$ to the probability at $V_{\mathrm{reset}}$ (eq.~(\ref{eq_fp})).

The description with the Fokker-Planck method is consistent with the LIF simulation in the limit of the number of neurons belonging to each sublattice $ND(\bm{\xi}) \to \infty$.  Because $\mathrm{MIN}(D(\bm{\xi})) \le 2^{-p}$, we restrict the total number of memory patterns to $p \sim O(1)$.

\section{Result}
\subsection{Activation of a single pattern}\label{sec_single}
We have previously reported that an input similar to a memory pattern cause the propagation of synchronous spikes of the memory pattern when the pattern rate $F=0.5$ in both of LIF simulation and the Fokker-Planck method~\cite{ishibashi}.  Here we answer whether the situation is same or not when the pattern rate $F \neq 0.5$.

The initial condition is a stationary distribution for no external input.  We activate the first layer of the network.  For the first layer activation, we consider the virtual layer 0 and describe the overlap on the virtual layer of the first memory pattern as a Gaussian function with standard deviation $\sigma$ and total volume $m^1$.  The volumes of the other memory patterns are set to 0.  Throughout this paper, the standard deviation $\sigma$ is always 0.5 ms.
\begin{align}
m^{0, \mu} (t)= \left\{
	\begin{array}{cl}
	\dfrac{m^1}{\sqrt{2 \pi} \sigma} \exp \left( \dfrac{(t-t_0)^2}{2 {\sigma}^2} \right) & \mu = 1,\\
	0 & \mu \neq 1.
	\end{array}
\right.\label{eq_input_single}
\end{align}

\begin{figure}[tb]
\centering
\includegraphics[width=7cm]{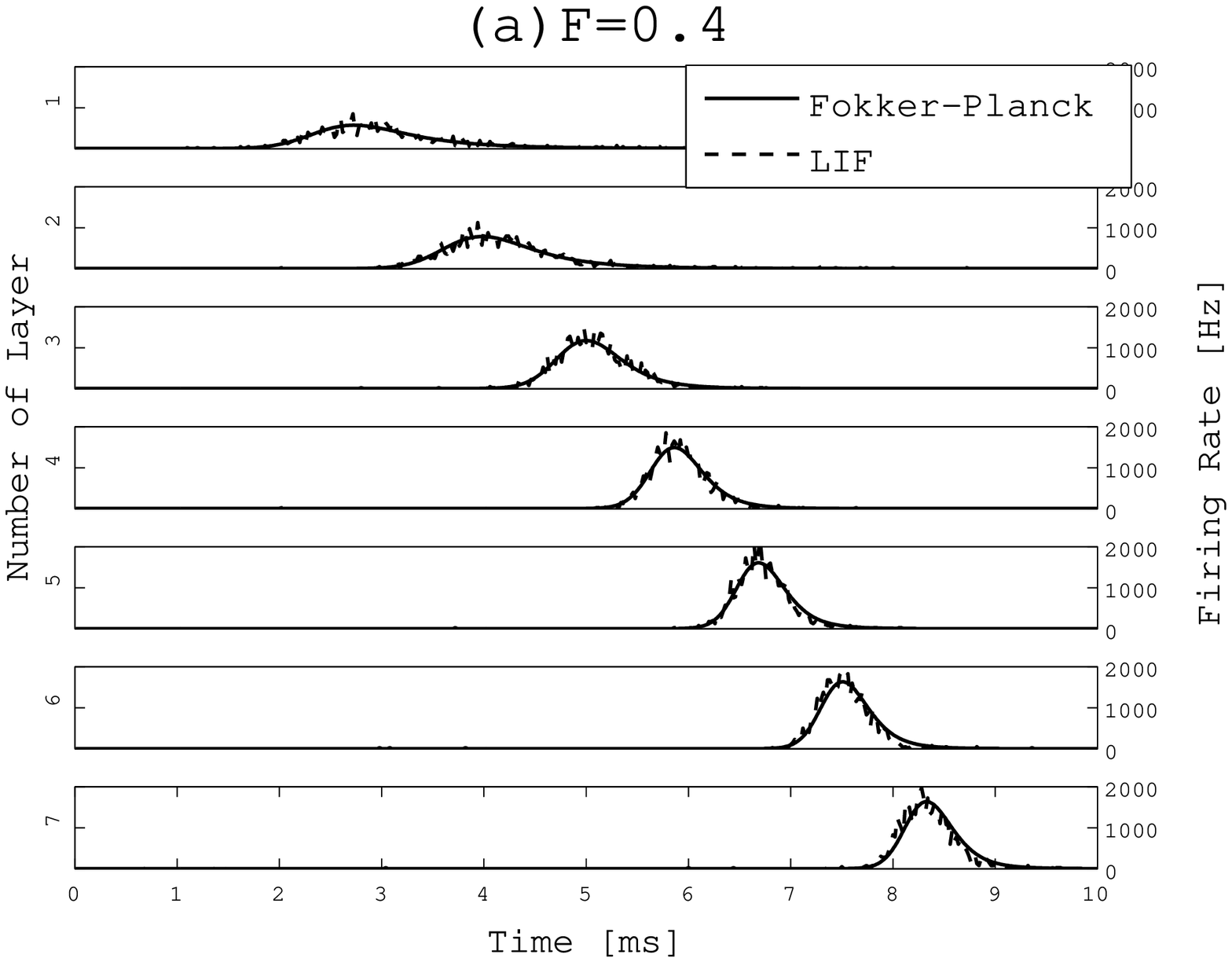}\\
\includegraphics[width=7cm]{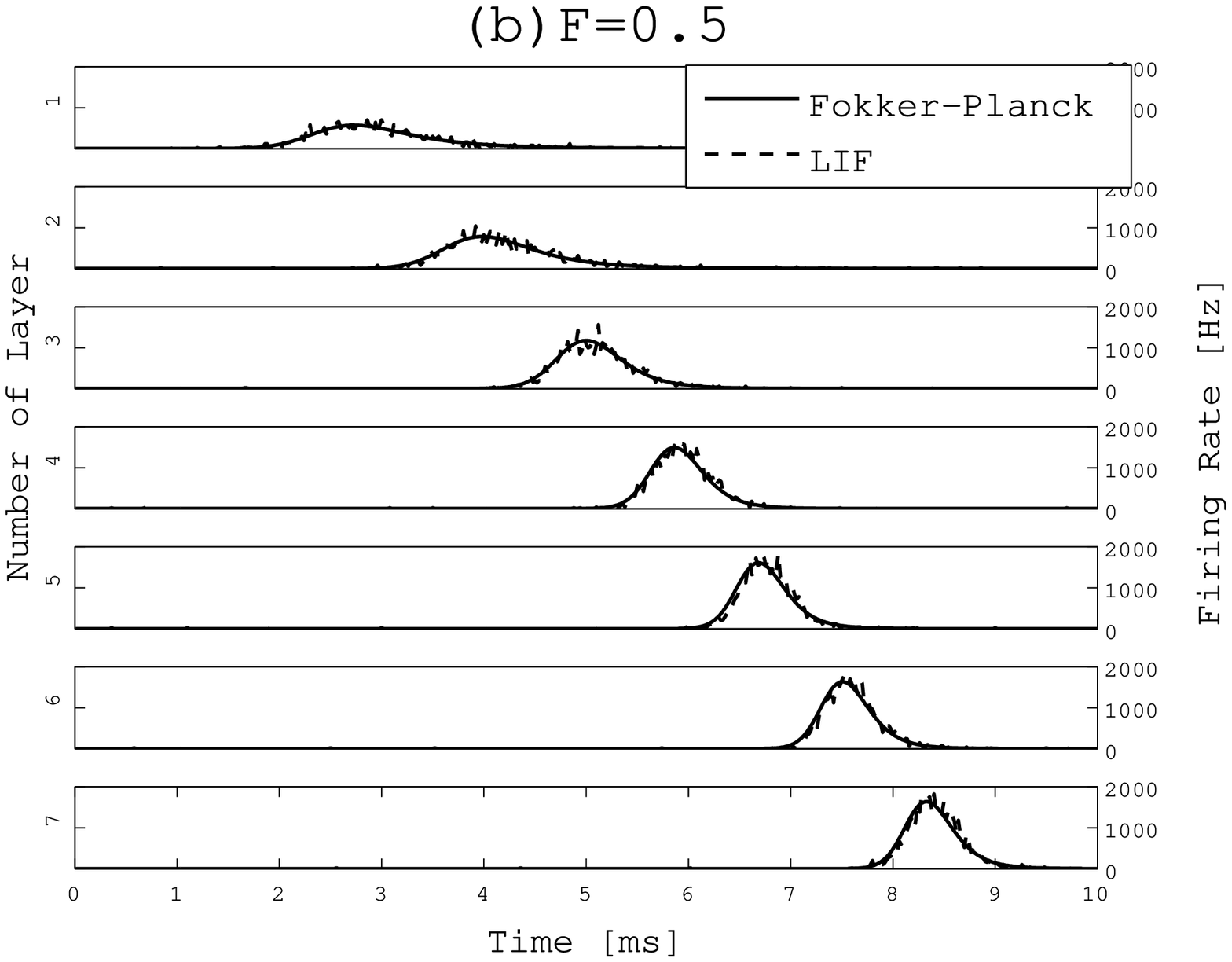}\\
\includegraphics[width=7cm]{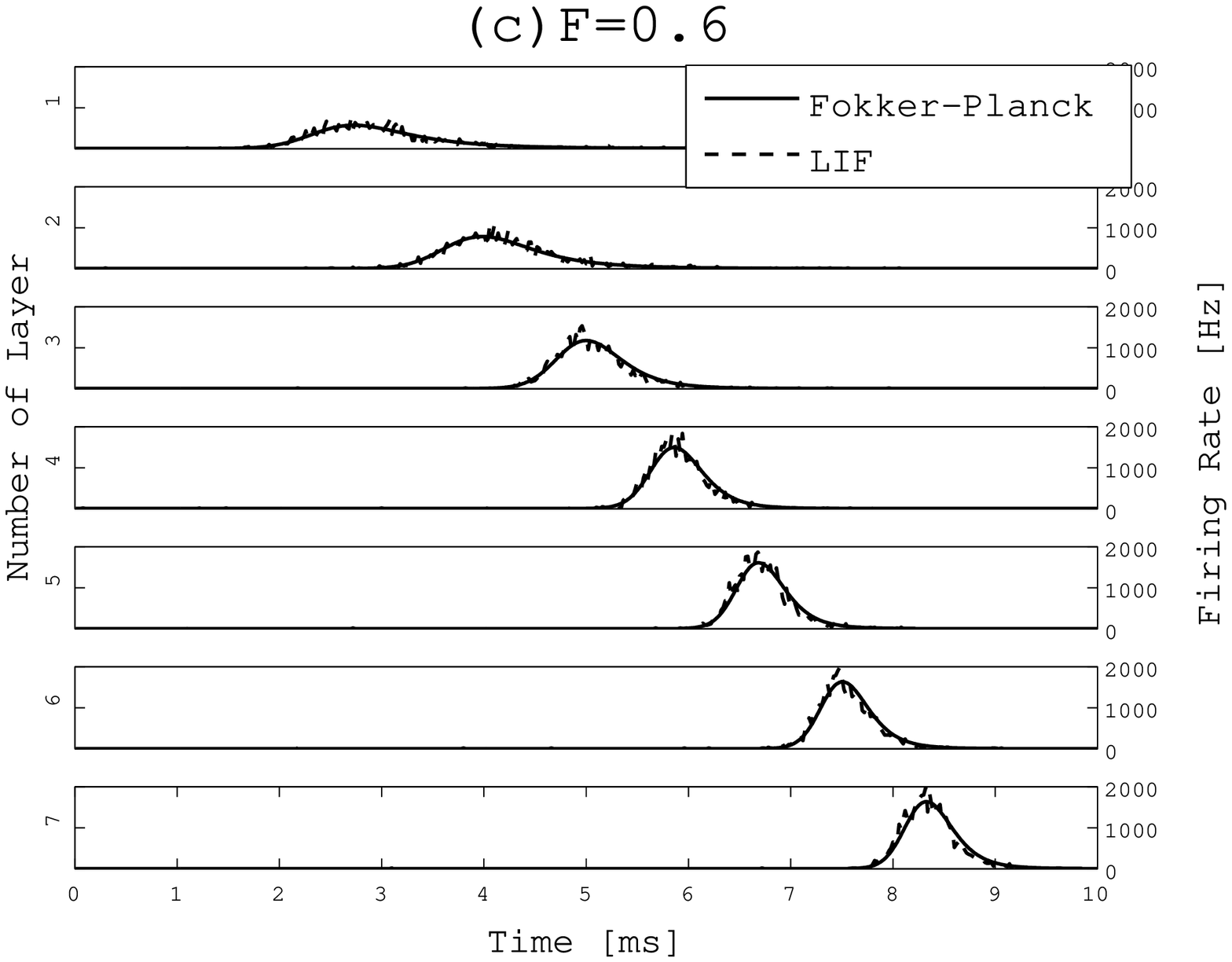}
\caption{Firing rates of the neurons whose first memory pattern $\xi^{l,1}_i$ is 1 on the vertical layer under the activation of a single pattern.  The pattern rate is $F=0.4$(a), $F=0.5$(b), and $F=0.6$(c).  The total volume of input $m^1$ is 0.6.  The dashed lines are obtained with the LIF simulation and the solid ones with the Fokker-Planck method.  We define $t^0$, when $m^{0,1}(t)$ takes a peak value, as $3\sigma=1.5$[ms].}
\label{fig_single}
\end{figure}

We calculate the input currents to neurons from eqs. (\ref{eq_ia}) and (\ref{eq_i}), membrane potential dynamics and firings from eq. (\ref{eq_lif}), and the overlaps from eq. (\ref{eq_m}).  Dashed lines in Fig. \ref{fig_single} show the firing rates of neurons whose first memory pattern $\xi^{l,1}_i$ is $+1$. Seven layers are vertically aligned from top to bottom.  Figure \ref{fig_single}(a) is the case of sparsely connected network $F=0.4$, Fig. \ref{fig_single}(b) is the case of conventional network $F=0.5$, and Fig. \ref{fig_single}(c) is the case of densely connected network $F=0.6$.  Synchronous spikes propagate in not only the case $F=0.5$ but also $F=0.4$ and 0.6.

Next, we apply the Fokker-Planck method.  We calculate the input current to the neurons belonging to each sublattice from eqs. (\ref{eq_ia_fp}) and (\ref{eq_i_fp}), the membrane potential distributions from eq. (\ref{eq_fp}), the firing rates from eq. (\ref{eq_j_fp}), and the overlaps from eq. (\ref{eq_m_fp}).  Since the overlaps of the $\mu (\neq 1)$th memory pattern are always 0, it is enough to divide the neurons into the sublattices only with the first memory pattern~\cite{ishibashi}.  Therefore, we consider two membrane distributions, $P^l_+(v,t)$ and $P^l_-(v,t)$, and two firing rates, $\nu^l_+(t)$ and $\nu^l_-(t)$ on each layer.  The subscript index $(+)$ means $\bm{\xi}=(+1)$ sublattice and $(-)$ means $\bm{\xi}=(0)$ sublattice.  Solid lines in Fig. \ref{fig_single} shows the firing rates of $\bm{\xi}=(+1)$, $\nu^l_+(t)$ on the seven layers vertically.  

Figure \ref{fig_single} suggests that regardless of pattern rate $F$ in the layered associative network the spikes of the input memory pattern synchronously propagate under a single pattern activation.  Figure~\ref{fig_single} also shows the consistency between the results of the LIF simulation and those of the Fokker-Planck method.
\subsection{Activation of two patterns}\label{sec_double}
We have previously reported that neurons belonging to different subllatices fire in different timing under the condition of activating two patterns~\cite{ishibashi}.  The spiking timing splits even when the input memory pattern is similar to one of the memory pattern; $m^1 \sim 1 and m^2 \sim 0$.  Here we address whether the spiking timing splits or not when the pattern rate $F \neq 0.5$.

Here we focus on the two memory patterns, and the other memory patterns have no overlaps.  Therefore we divide neurons into sublattices according to the signs of the two patterns as well as \S \ref{sec_single}.  The sublattices $\bm{\xi}=(+1,+1)$, $(+1,0)$, $(0, +1)$, and $(0,0)$ are respectively described as $(++)$, $(+-)$, $(-+)$, and $(--)$.  We accordingly denote the firing rates of each sublattice as $\nu^l_{++}(t)$, $\nu^l_{+-}(t)$, $\nu^l_{-+}(t)$, and $\nu^l_{--}(t)$.

Throughout \S \ref{sec_double}, we consider the following overlaps of the virtual layer.
\begin{align}
m^{0, 1} (t)=& \dfrac{1}{\sqrt{2 \pi} \sigma} \biggl( I^0_{++}F \exp \left( \dfrac{(t-t_{++})^2}{2 {\sigma}^2} \right) \nonumber \\
& \quad + I^0_{+-}(1-F) \exp \left( \dfrac{(t-t_{+-})^2}{2 {\sigma}^2} \right) \biggr), \\
m^{0, 2} (t)=& \dfrac{1-F}{\sqrt{2 \pi} \sigma} \biggl( I^0_{++}\exp \left( \dfrac{(t-t_{++})^2}{2 {\sigma}^2} \right) \nonumber \\
& \quad  - I^0_{+-}\exp \left( \dfrac{(t-t_{+-})^2}{2 {\sigma}^2} \right) \biggr), \\
m^{0, \mu} (t)=& 0  \qquad \qquad \mu \neq 1,2.
\end{align}
Then from eq. (\ref{eq_i_fp}) the input to $(++)$ and $(+-)$ sublattices on the first layer is simply described as follows,
\begin{align}
I^1_{++}(t) = \dfrac{I^0_{++}}{\sqrt{2 \pi} \sigma} \exp \left( \dfrac{(t-t_{++})^2}{2 {\sigma}^2} \right), \label{eq_input1}\\
I^1_{+-}(t) = \dfrac{I^0_{--}}{\sqrt{2 \pi} \sigma} \exp \left( \dfrac{(t-t_{+-})^2}{2 {\sigma}^2} \right). \label{eq_input2}
\end{align}

In \S \ref{sec_double} the results of the LIF simulation are not shown but we confirmed that their results are consistent with those of the Fokker-Planck method.
\subsubsection{Different Strength of Input}\label{sec_diffstr}
In \S \ref{sec_diffstr} we consider the situation that the volume of the overlaps of the first and second memory pattern on the virtual layer are respectively set to $m^1$ and $m^2$, and the timing of input to $(++)$ and $(+-)$ is set to the same; $I^0_{++}F+I^0_{+-}(1-F)=m^1$, $(1-F)(I^0_{++}-I^0_{+-})=m^2$ and $t_{++}=t_{+-}=t_0$ in eqs. (\ref{eq_input1}) and (\ref{eq_input2}).  Therefore eqs. (\ref{eq_input1}) and (\ref{eq_input2}) are rewritten as
\begin{align}
I^1_{++}(t) = \dfrac{m^1+m^2}{\sqrt{2 \pi} \sigma} \exp \left( \dfrac{(t-t_0)^2}{2 {\sigma}^2} \right),\\
I^1_{+-}(t) = \dfrac{(1-F)m^1-F m^2}{(1-F)\sqrt{2 \pi} \sigma} \exp \left( \dfrac{(t-t_0)^2}{2 {\sigma}^2} \right).
\end{align}

\begin{figure}[t]
\centering
\includegraphics[width=7cm]{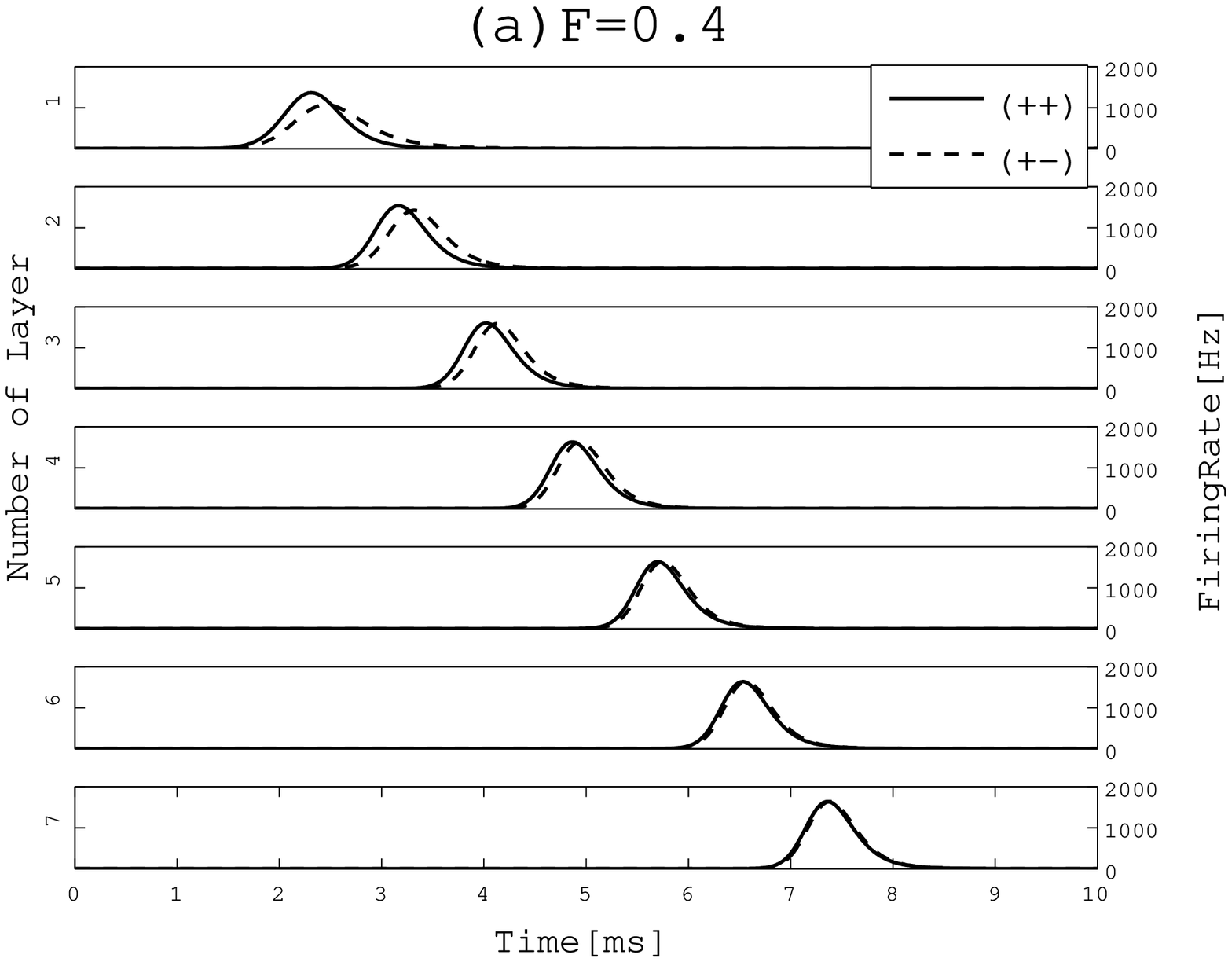}\\
\includegraphics[width=7cm]{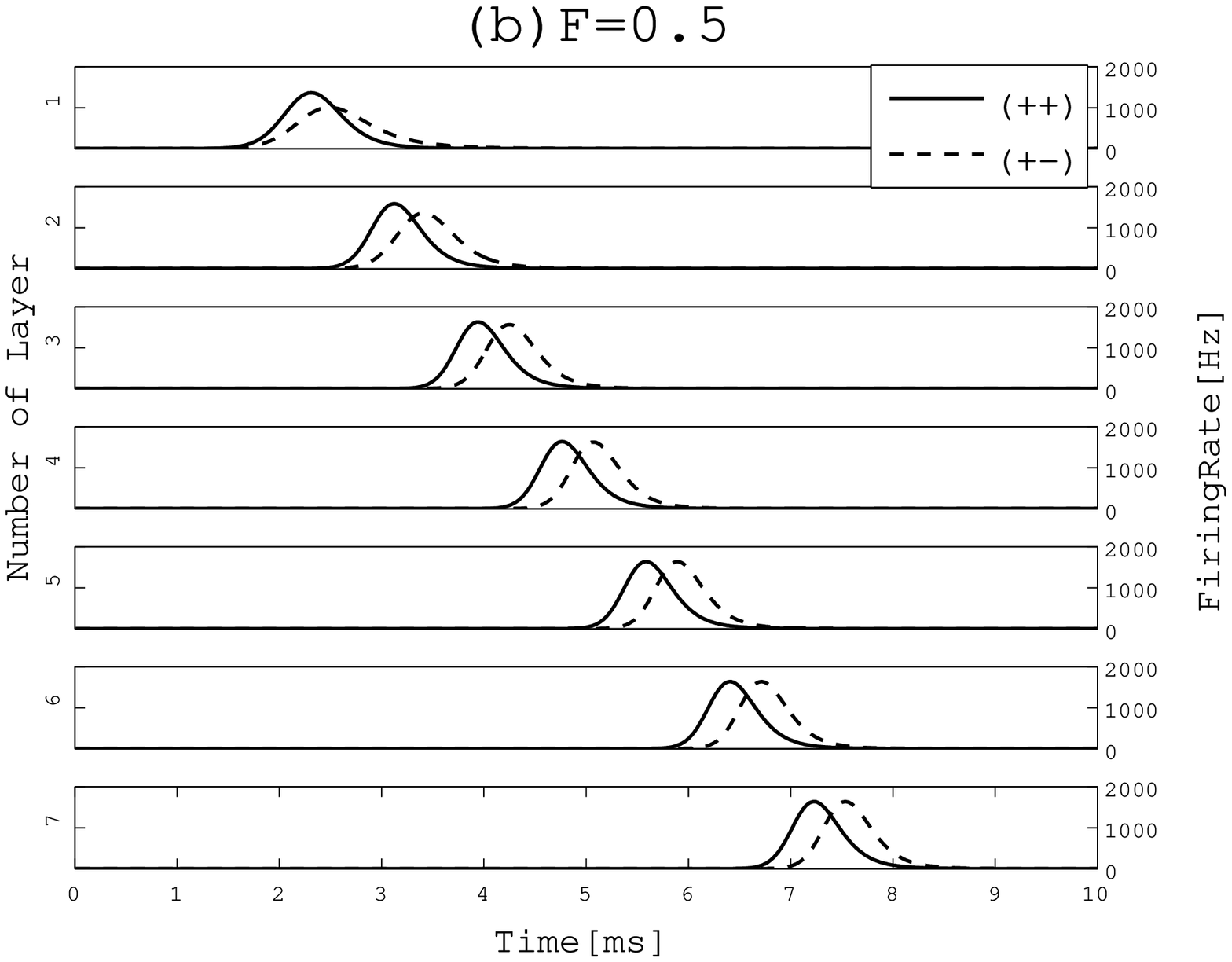}\\
\includegraphics[width=7cm]{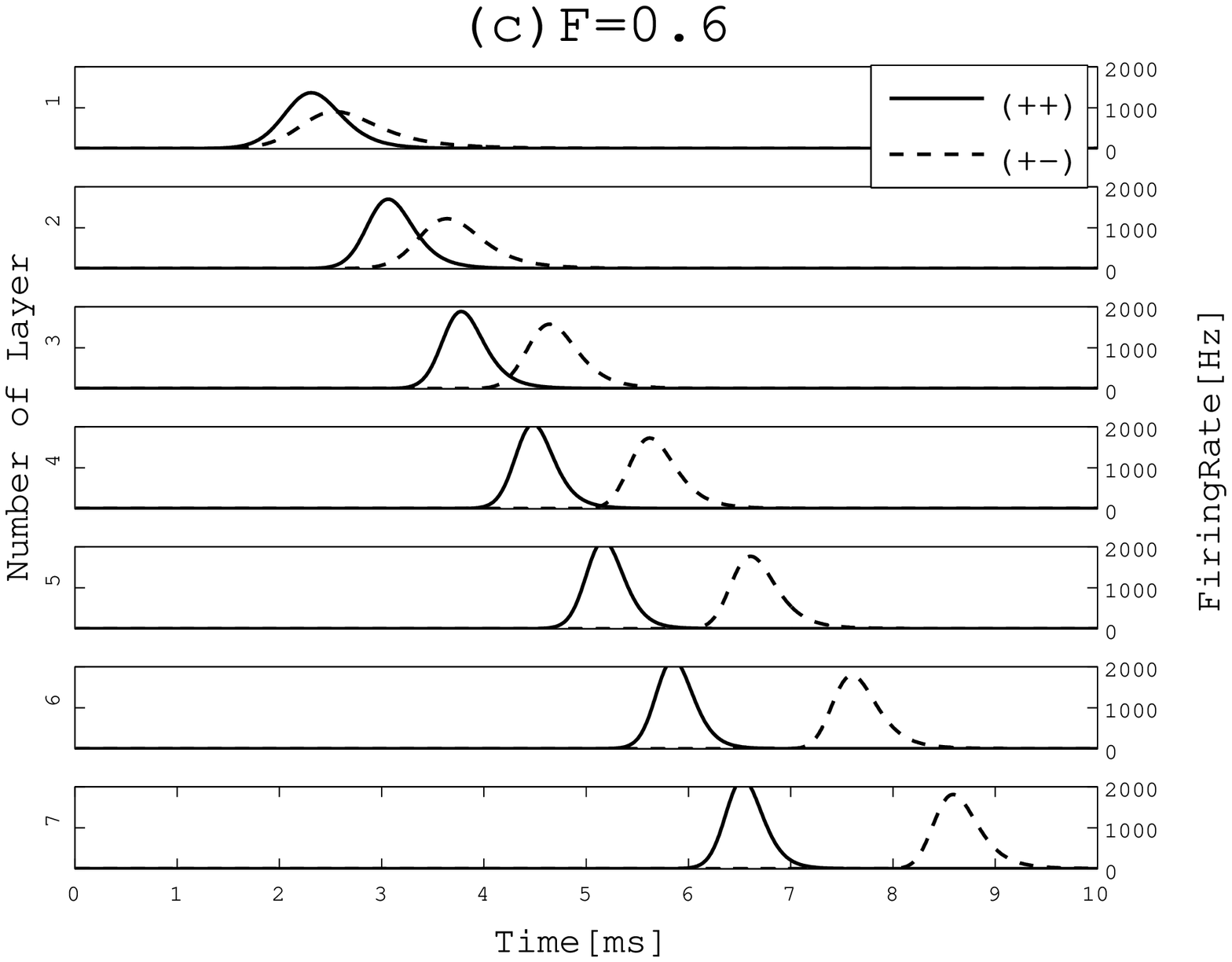}
\caption{
Firing rates of $(++)$ sublattices ($\nu^l_{++}(t)$; solid lines) and $(+-)$ ones ($\nu^l_{+-}(t)$; dashed lines) on the vertical layer when the strength of activation is different.  The pattern rate is $F=0.4$(a), $F=0.5$(b), and $F=0.6$(c).  The total volume of input of the first memory pattern $m^1$ is 0.9 and that of the second memory pattern $m^2$ is 0.1.  We define $t^0$, when $m^{0,1}(t)$ takes a peak value, as $3\sigma=1.5$[ms].  These results is obtained with the Fokker-Planck method.
}
\label{fig_diffstr}
\end{figure}

Here we focus on the case that the input is similar to the first memory pattern; $m^1 \sim 1$ and $m^2 \sim 0$.  If $m^2=0$, the input to $(++)$ and $(+-)$ sublattices is same, $I^l_{++}(t) = I^l_{+-}$ and then the first memory pattern propagates in the shape of synchronous firing packet as shown is \S \ref{sec_single} (Fig. \ref{fig_single}).  We calculate in the case of $m^1=0.9$, $m^2=0.1$ by using the Fokker-Planck method.  Figure~\ref{fig_diffstr} shows the firing rates $\nu^l_{\bm{\xi}}(t)$.  Solid lines indicate the firing rates of $(++)$ sublattices $\nu^l_{++}(t)$ and dashed ones are for the firing rates of $(+-)$ sublattices $\nu^l_{+-}(t)$.  The pattern rate is $F=0.4$ (Fig. \ref{fig_diffstr}(a)), $F=0.5$ (Fig. \ref{fig_diffstr}(b)), and $F=0.6$ (Fig. \ref{fig_diffstr}(c)).  When the pattern rate $F$ is 0.5 (Fig. \ref{fig_diffstr}(b)), the spikes of $(++)$ and $(+-)$ sublattices propagates in different timing, as previously reported~\cite{ishibashi}.  When the pattern rate $F$ is 0.4 (Fig. \ref{fig_diffstr}(a)), at the beginning, the neurons of $(++)$ and $(+-)$ sublattices fires in different timing but after propagation of several layers they become to fire synchronously.  On the contrary, when the pattern rate $F$ is 0.6, the timing difference between $(++)$ and $(+-)$ sublattices becomes larger as spikes propagate.

These results imply that in the network of the pattern rate $F<0.5$, that is the sparsely connected network, synchronous firing between sublattices is promoted and in that of $F>0.5$, that is the densely connected network, synchronous firing is suppressed.

The cause of these promotion and suppression of synchronous firing can be understood from the input currents to each sublattice $I^l_{\bm{\xi}}(t)$.  The input currents are described as follows:
\begin{align}
I^{l+1}_{++} &= m^{1,l}(t) + m^{2,l}(t) \nonumber \\
&= 2F\nu^l_{++}(t)  - 2(1-F)\nu^l_{--}(t) \nonumber\\
&\qquad+ (1-2F) (\nu^l_{+-}(t) + \nu^l_{-+}(t) ),\label{eq_int1}\\
I^{l+1}_{+-} &= m^{1,l}(t) - \frac{F}{1-F}m^{2,l}(t) \nonumber \\
&= \frac{1-2F}{1-F} \left( F \nu^l_{++}(t) - (1-F) \nu^l_{--}(t) \right) \nonumber\\
&\qquad+ \frac{2F^2-2F+1}{1-F}\nu^l_{+-}(t) - 2F \nu^l_{-+}(t).\label{eq_int2}
\end{align}
When the pattern rate $F=0.5$, that is $1-2F=0$, $(++)$ sublattices do not interact with $(+-)$.  Therefore the timing difference caused by the difference of input strength does not change during propagation~\cite{ishibashi}.  When the pattern rate $F$ is less than 0.5, that is $1-2F$ has positive value, there are excitatory connections from $(++)$ and $(+-)$ sublattices to $(+-)$ and $(++)$ ones on the next layer respectively.  The excitatory connections seem to promote the synchronous firing like a synfire chain.  That is why the timing difference decreases as spikes propagate.  On the contrary, when the pattern rate $F$ is more than 0.5, that is $1-2F$ has negative value, there are inhibitory connections as well.  The inhibitory connection seems to suppress the synchronous firing and made the timing difference larger as spikes propagate.

\subsubsection{Different Timing of Input}\label{sec_difftime}
In \S \ref{sec_diffstr} when the strength of input to sublattices is different, it seems that sparsely and densely connected network respectively promote and suppress synchronous firing.  Here we observe the activity when we set a difference not in the strength but in the timing of input to sublattices; $I^0_{++}=I^0_{+-}=I^0$ in eqs. (\ref{eq_input1}) and (\ref{eq_input2}).  Then eqs. (\ref{eq_input1}) and (\ref{eq_input2}) are rewritten as
\begin{align}
I^1_{++}(t) = \dfrac{I^0}{\sqrt{2 \pi} \sigma} \exp \left( \dfrac{(t-t_{++})^2}{2 {\sigma}^2} \right), \\
I^1_{+-}(t) = \dfrac{I^0}{\sqrt{2 \pi} \sigma} \exp \left( \dfrac{(t-t_{+-})^2}{2 {\sigma}^2} \right).
\end{align}

We calculate in the case that the input to $(++)$ is earlier than that of $(+-)$ by 1ms, that is $t_{+-}=t_{++}+1$[ms].

\begin{figure}[t]
\centering
\includegraphics[width=7cm]{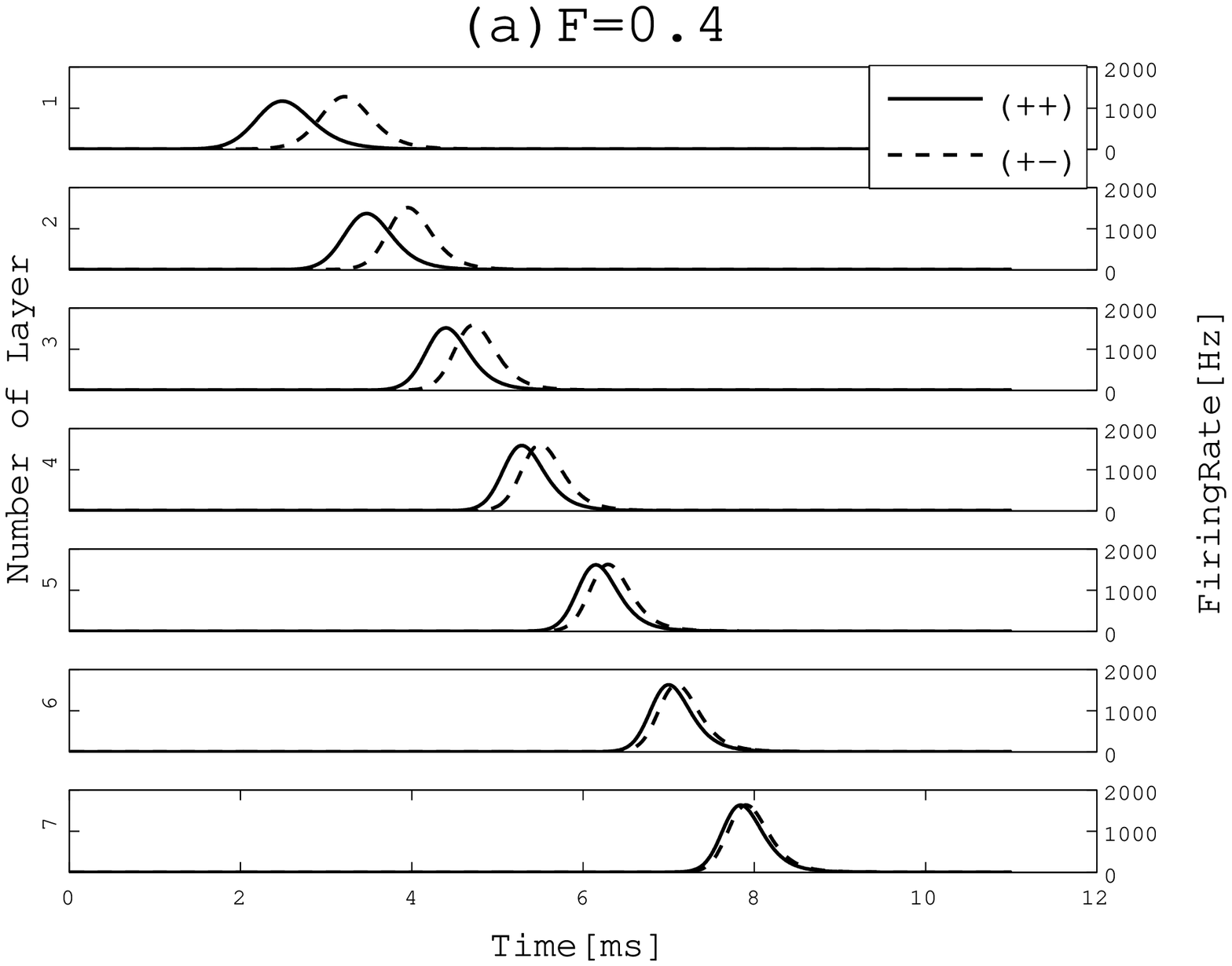}\\
\includegraphics[width=7cm]{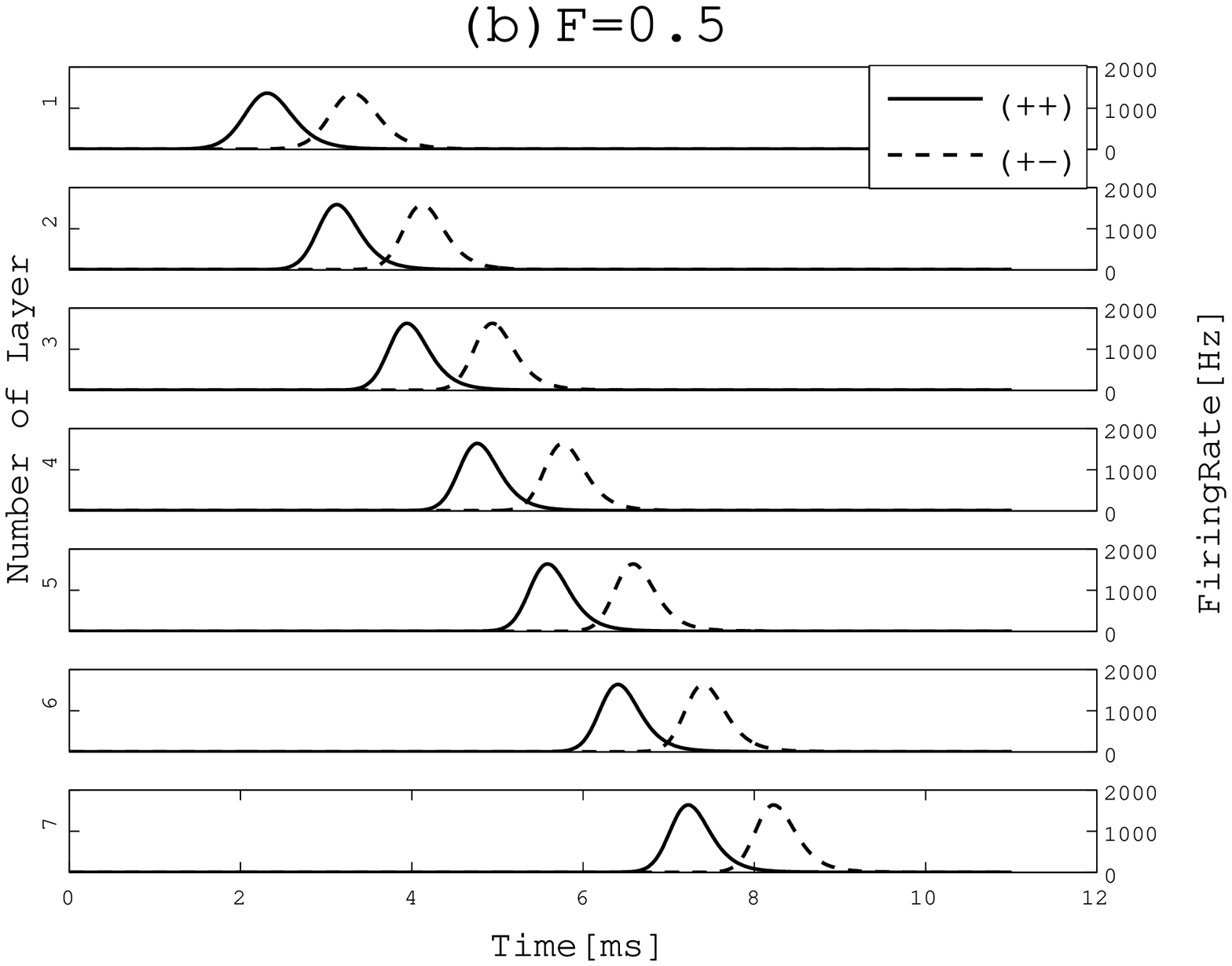}\\
\includegraphics[width=7cm]{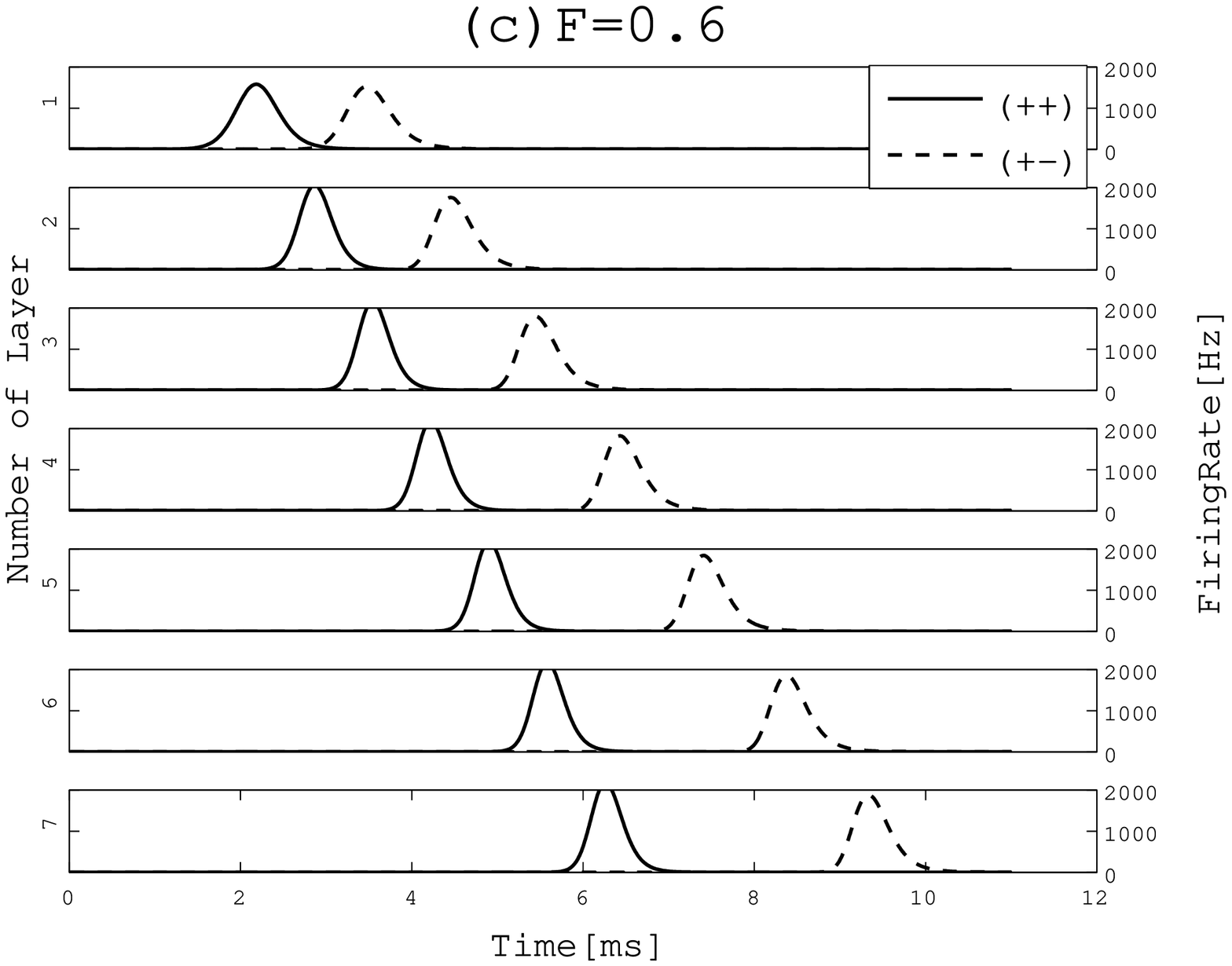}
\caption{
Firing rates of $(++)$ sublattices ($\nu^l_{++}(t)$; solid lines) and $(+-)$ ones ($\nu^l_{+-}(t)$; dashed lines) on the vertical layer when the timing of activation is different.  The pattern rate is $F=0.4$(a), $F=0.5$(b), and $F=0.6$(c).  The size of input $I^0$ is 1.  We define $t_{++}$, when $I^0_{++}(t)$ takes a peak value, as $3\sigma=1.5$[ms] and $t_{+-} = t_{++}+1$[ms]$=$2.5[ms].  These results is obtained with the Fokker-Planck method.
}
\label{fig_difftime}
\end{figure}

Figure~\ref{fig_difftime} shows the firing rates $\nu^l_{\bm{\xi}}(t)$.  Solid lines indicate the firing rates of $(++)$ sublattices $\nu^l_{++}(t)$ and dashed ones are for the firing rates of $(+-)$ sublattices $\nu^l_{+-}(t)$ on the seven layers.  When the pattern rate $F$ is 0.5 (Fig. \ref{fig_difftime}(b)), the spikes of $(++)$ and $(+-)$ propagates at the same speed.  When the pattern rate $F$ is 0.4 (Fig. \ref{fig_difftime}(a)) the timing difference becomes smaller, and when the pattern rate $F$ is 0.6 the timing difference becomes larger as spikes propagate.  All the results of Figs.~\ref{fig_difftime}(a-c) is consistent with those of \S \ref{sec_diffstr}

\subsubsection{Basin of Attraction}\label{sec_attract}
\begin{figure}[t]
\centering
\includegraphics[width=6cm]{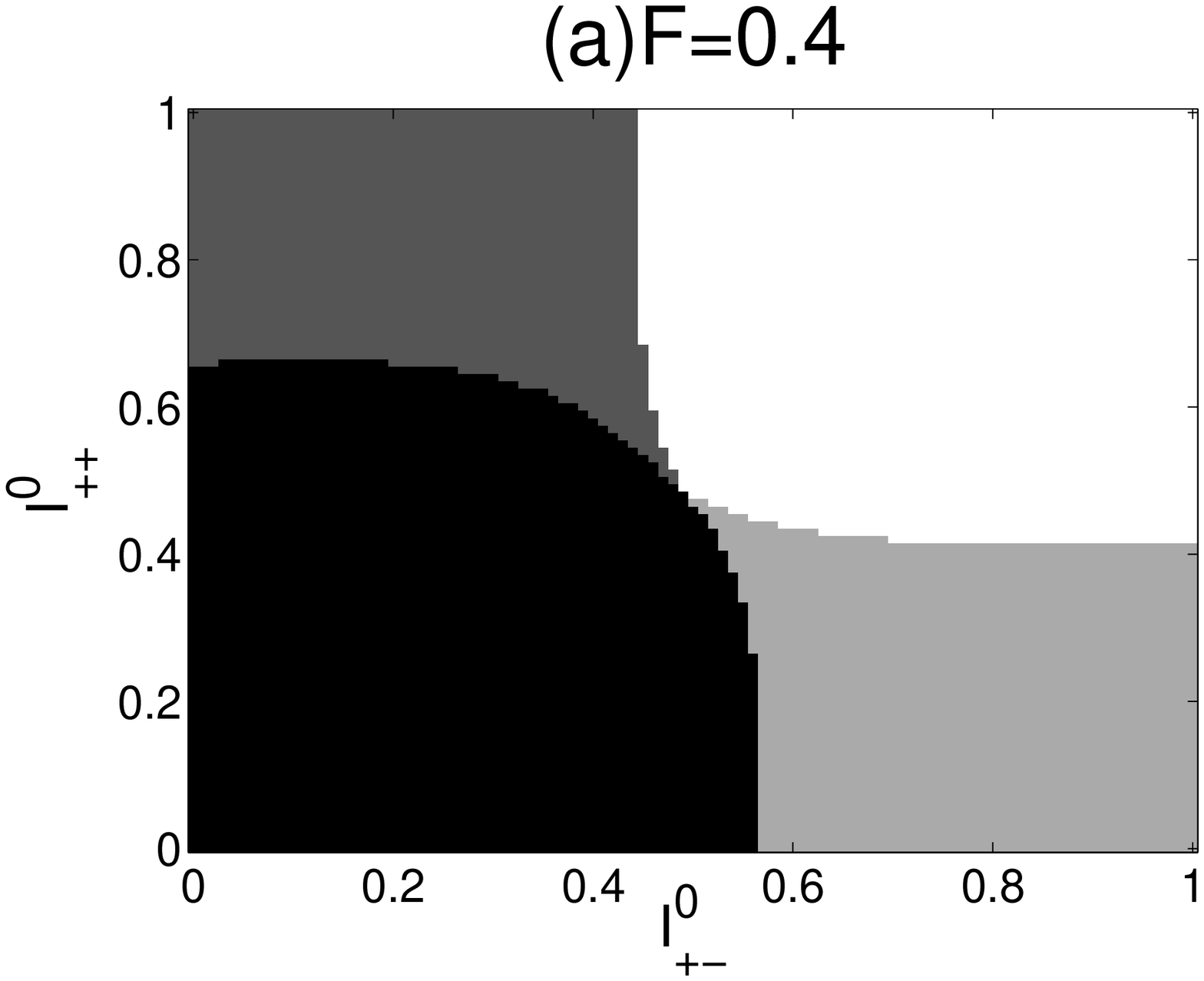}
\includegraphics[width=6cm]{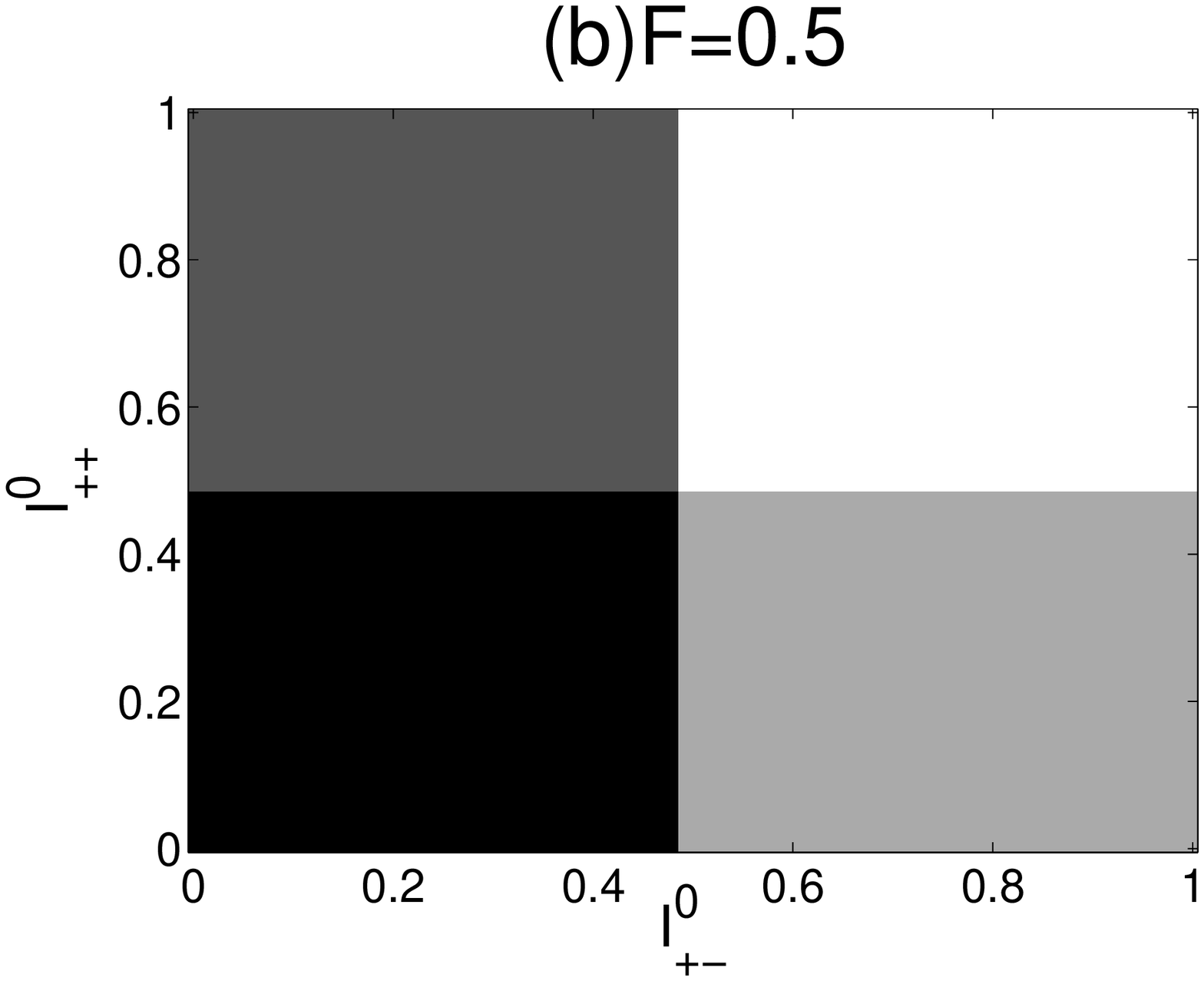}
\includegraphics[width=6cm]{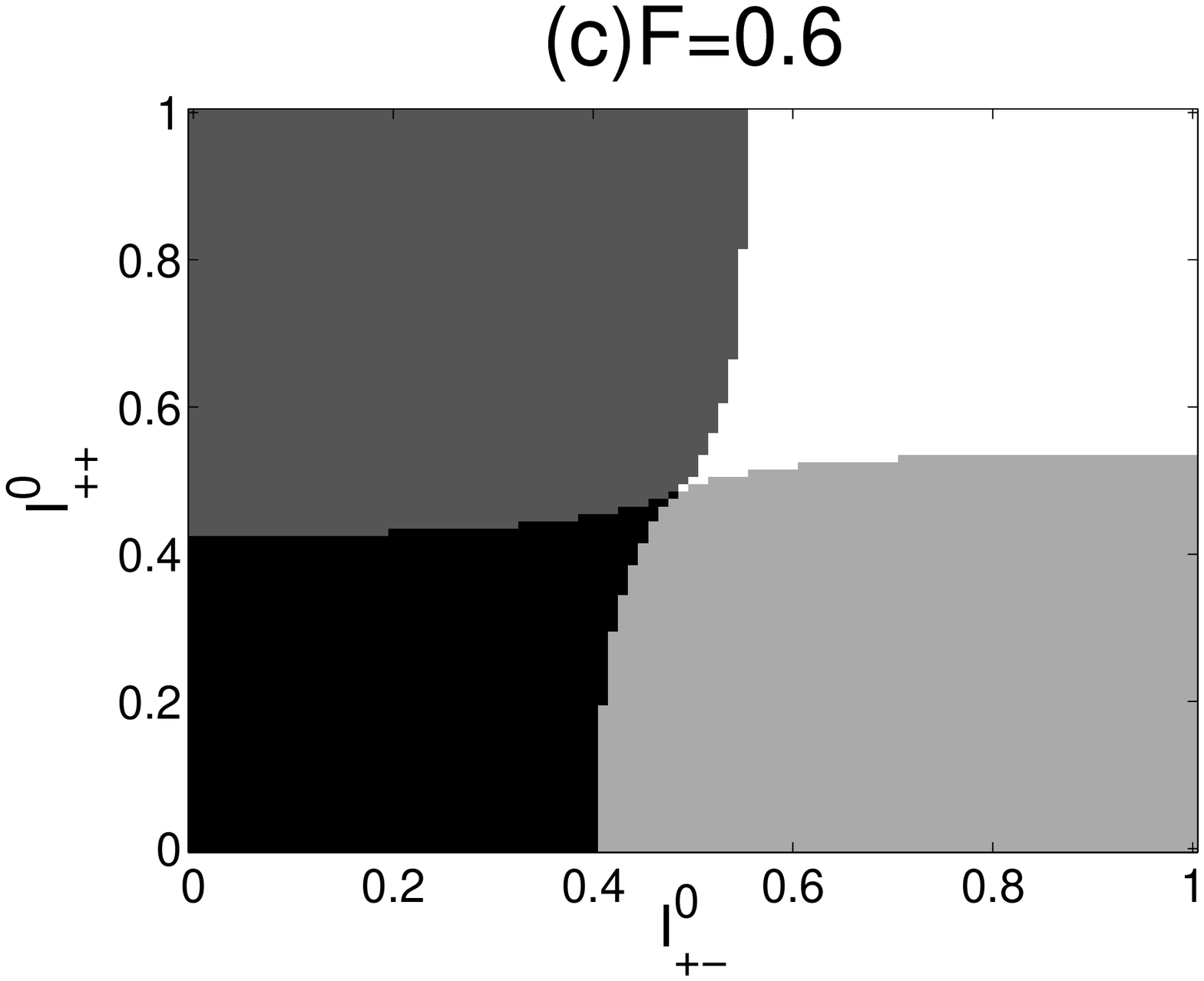}
\caption{
The firing of each sublattices.  The pattern rate is $F=0.4$(a), $F=0.5$(b), and $F=0.6$(c).  The vertical axis means the input strength of $(++)$ $I^0_{++}$ and the horizontal axis means that of $(+-)$ $I^0_{+-}$.  If the maximum of the firing rate is more than 600[Hz] on the fifth layer, we regard the sublattice as 'firing'.  The black, dark gray, light gray and white region in respectively mean no firing, activity in $(++)$ sublattice is propagated, activity in $(+-)$ sublattice is propagated, and the first memory pattern is fully associated and activity in both $(++)$ and $(+-)$ sublattices are propagated.
}
\label{fig_attract}
\end{figure}

In \S \ref{sec_diffstr} and \S \ref{sec_difftime}, we have shown that the sparsely and densely connected network seems to respectively promote and suppress synchronous firing between sublattices.  Here we focus on not firing timing but the stability of firing of sublattices.  The timing of inputs to $(++)$ and $(+-)$ is set to be same, i.e., $t_{++}=t_{+-}=t_0$ in eqs. (\ref{eq_input1}) and (\ref{eq_input2}).  Then eqs. (\ref{eq_input1}) and (\ref{eq_input2}) are rewritten as
\begin{align}
I^1_{++}(t) = \dfrac{I^0_{++}}{\sqrt{2 \pi} \sigma} \exp \left( \dfrac{(t-t_0)^2}{2 {\sigma}^2} \right),\\
I^1_{+-}(t) = \dfrac{I^0_{+-}}{\sqrt{2 \pi} \sigma} \exp \left( \dfrac{(t-t_0)^2}{2 {\sigma}^2} \right).
\end{align}
We observe the firing rates when we change the input strength of $(++)$ subllatices $I^0_{++}$ and $(+-)$ one $I^0_{+-}$ independently from 0 to 1.  If the maximum of the firing rate of a sublattice on the fifth layer is more than 600[Hz], we regard the sublattice fires.  Figure~\ref{fig_attract} is the result obtained with the Fokker-Planck method.  The vertical axis means the input strength of $(++)$ $I^0_{++}$ and the horizontal axis means that of $(+-)$ $I^0_{+-}$.  The black, dark gray, light gray and white region in Fig.~\ref{fig_attract} respectively mean no firing, activity in $(++)$ sublattice is propagated, activity in $(+-)$ sublattice is propagated, and the first memory pattern is fully associated and activity in both $(++)$ and $(+-)$ sublattices are propagated.  The pattern rate is $F=0.4$(a), $F=0.5$(b), and $F=0.6$(c).  When the pattern rate $F=0.4$ (Fig.~\ref{fig_attract}(a)) the region of the first memory pattern is larger than that in the case of $F=0.5$ (Fig.~\ref{fig_attract}(b)).  On the contrary, when the pattern rate $F=0.6$ (Fig.~\ref{fig_attract}(c)) the region of the first memory pattern is smaller than that in the case of $F=0.5$ (Fig.~\ref{fig_attract}(b)).

These results imply that the sparsely and densely connected network not only promotes and suppresses synchronous firing but also enlarges and shrinks the basin of attraction of the memory pattern respectively.  The cause of enlargement and shrinkage seems to be the excitatory and inhibitory connections between sublattices as described in \S \ref{sec_diffstr}.  Under the existence of excitatory connections between $(++)$ and $(+-)$ sublattices, $(++)$ and $(+-)$ sublattices mutually excite each other.  On the other hand, under the existence of inhibitory connections, $(++)$ and $(+-)$ sublattices mutually inhibit each other.

\subsection{Storage Capacity}\label{sec_cap}
In the binary neurons network it has reported that sparse connection increases the storage capacity of memory patterns~\cite{amari89, amari91,okada}.  Here we show the results in the case of the LIF neurons.  The number of neurons per layer $N$ is set to 5000, and we change the total number of the memory pattern $p$ from 1 to 500.  The input is written by eq. (\ref{eq_input_single}) as well as \S \ref{sec_single}, but the total volume of the input $m^1=1$.

\begin{figure}[t]
\centering
\includegraphics[width=7cm]{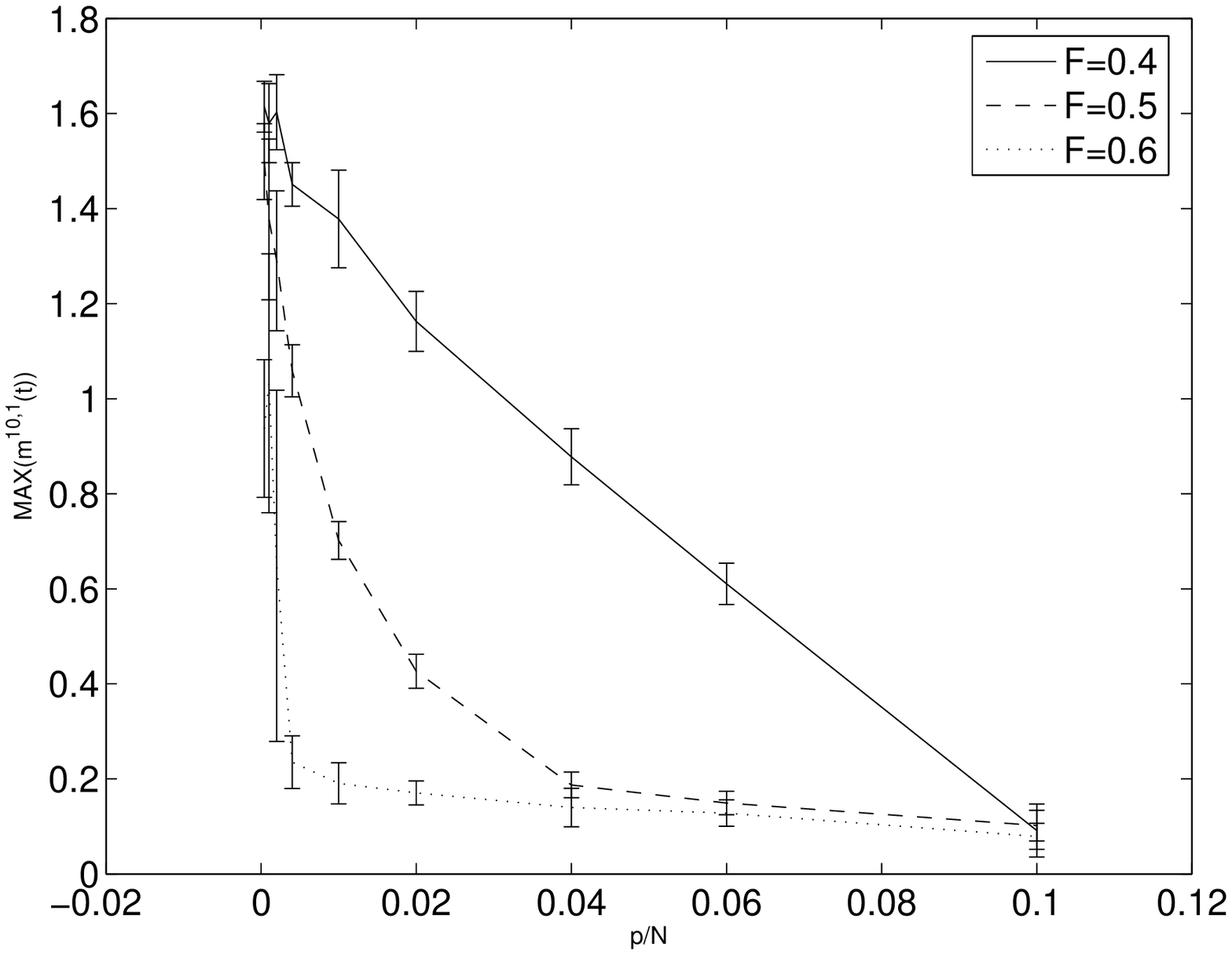}
\caption{
The maximum value of the overlap of the input memory pattern on the 20th layer.  The horizontal axis means the rate of the total number of memory pattern $p$ to the number of the neurons per layer $N$.  We averaged the overlap over 10 trials and shows the standard deviations as the error-bars.
}
\label{fig_cap}
\end{figure}

Figure \ref{fig_cap} shows the maximum value of the overlap of the input memory pattern on the 20th layer.  This figure suggests that the smaller the pattern rate $F$, the more stable the propagating patterns are.  It seems that the result is also caused by the excitatory and inhibitory connections because synchronous firing enlarges the maximum value of the overlap.

\section{Summary and Discussion}
In this paper, we studied the activity of a layered associative network constructed by the LIF neurons with taking into account of the sparseness of the memory patterns.  The effect of sparseness has been mainly studied in recurrent networks~\cite{amari89, amari91, okada}.  In the layered network with spiking neurons, memory patterns propagate in the shape of synchronized pulse packet.  The sparseness increases the storage capacity (\S \ref{sec_cap}), which coincide with the result of the recurrent networks.  The sparseness also affects the propagation of synchronous pulse packets between sublattices in the feedforward network case.  In two patterns activation sparse(dense) connection promotes(suppresses) the propagation of synchronous pulse packets between sublattices (\S \ref{sec_double}) in the feedforward network of spiking neurons.

The increase of storage capacity imply the superiority of the sparse connection in the feedforward associative networks.  On the other hands, the role of synchronous firing observed in sparsely connected networks remains to be elucidated.  Future studies will be to elucidate how the neural networks can use such synchronous propagation of pulse packets in the information processing.

\section*{Acknowledgments}
This work was partially supported by a Grant-in-Aid for Scientific Research on Priority Areas No. 14084212, and for Scientific Research (C) No. 16500093 from the Ministry of Education, Culture, Sports, Science and Technology of Japan.

\end{document}